\documentclass[12pt,preprint]{aastex}

\slugcomment{Accepted by The Astronomical Journal} 

\shorttitle{Compact Radio Sources in Orion}
\shortauthors{Zapata, Rodr\'\i guez, Kurtz, \& O'Dell}

\begin{document}

\title{Compact Radio Sources in Orion: New Detections, Time Variability, and Objects in
OMC-1S}

\author{Luis A. Zapata, Luis F. Rodr\'\i guez, and Stanley E. Kurtz} 
\affil{Centro de Radioastronom\'\i a y Astrof\'\i sica, UNAM, 
Apdo. Postal 3-72 (Xangari), 58089 Morelia, Michoac\'an, M\'exico}
\email{l.zapata, l.rodriguez, s.kurtz@astrosmo.unam.mx}

\and

\author{C. R. O'Dell}
\affil{Department of Physics and Astronomy, Vanderbilt University, Box 1807-B, 
Nashville, TN 37235}
\email{cr.odell@vanderbilt.edu}

\begin{abstract}

We present the analysis of four 3.6~cm radio continuum archival observations
of Orion obtained
using the Very Large Array in its A-configuration,
with $0\rlap.{''}3$ angular resolution. The observations were made 
during the period 1994-1997.
In a region of $4' \times 4'$, we detect a total
of 77 compact radio sources. Of the total of
detected sources, 54 are
detected in one or more of the individual
observations and 36 of these show time variability (by more than
30\%) between the observed epochs. 
A deep image made from averaging all data shows an additional 23 faint sources,
in the range of 0.1 to 0.3 mJy.
Of the total of 77 sources, 39 are new centimeter detections.
However, only 9 of the 77 sources do not have a previously
reported counterpart at
near-infrared, optical, or X-ray wavelengths.
In particular, we detect three faint sources in the OMC-1S region
that may be related to the sources that power the 
multiple outflows that emanate from
this part of the Orion nebula.

\end{abstract}  


\section{INTRODUCTION}

The Orion Nebula (NGC 1976, M42) is the closest region of massive star 
formation, 
and it is associated with a variety of compact
objects that include stars, nebulae, and masers, 
as well as
infrared, optical and X-rays sources (Gaume et al. 1998,  Ali \& Depoy 
1995, O'Dell \& Wen 1994, and Garmire et al. 2000). It is the best studied
region of its class.
A recent summary of the properties of Orion 
is given by O'Dell (2001).

The early identification by Laques \& Vidal (1979) of six unresolved 
objects near $\theta^{1}$
Ori C, determined to be high ionization emission line sources, and 
subsequent studies in the 
radio by Garay et al. (1987), Churchwell et al. (1987), and Felli et 
al. (1993a; 1993b), among others, suggested that possibly these objects
were compact neutral 
clouds photoionized externally by $\theta^{1}$ Ori C. 
The observations of these 
and similar sources in Orion
with the HST by O'Dell et al. (1993), established
their nature as externally ionized proto-planetary disks (proplyds).
We refer the
reader to Henney \& Arthur (1998) for a detailed description of the propylds.

In this paper we present an analysis of radio continuum VLA archival
observations of Orion, made at 3.6 cm in the highest angular
resolution configuration A.
One of the purposes of our analysis was to produce a deeper image that
could reveal additional faint radio sources to those previously reported
in the literature.
Indeed, we detected 77 compact radio sources, 39 of which are new
centimeter detections.

A region in Orion that has received considerable attention in recent
years is OMC-1S (Keene, Hildebrand,
\& Whitcomb 1982; Bally, O'Dell \& McCaughrean 2000), because 
it shows remarkable outflow activity.
However, the identification of the exciting sources of these outflows
remains unclear. We have used our deep, averaged image to search for continuum
sources in this region.

\section{OBSERVATIONS}

We have used 3.6 cm archival data from
the Very Large Array of the NRAO\footnote{The National Radio Astronomy
Observatory is a facility of the National Science Foundation operated
under cooperative agreement by Associated Universities, Inc.} 
in its A-array configuration to study the cluster of radio sources
in Orion and to search for new, fainter objects in a sensitive image
made from concatenating all the data.
The region was observed on
1994 April 29, 1995 July 22, 1996 November 21, and 1997 January 11.
The amplitude calibrator was
1331+305, with an adopted flux density of 5.18 Jy,
and the phase calibrator was 0501$-$019, for all epochs except
1995, when 0541$-$056 was used. The HPFW of the primary
beam was 6$'$. In Table 1 we present a summary of
the observations.

The data were analyzed in the standard manner
using the package AIPS of NRAO. The data were also self-calibrated in phase
and amplitude for each epoch. 
Individual images were made at each epoch to search for variability
between the epochs observed.
To diminish the effects of extended emission, 
we used only visibilities
with baselines longer than 100 k$\lambda$, thus suppressing the emission
of structures larger than $\sim$2${''}$.
Images were made with the ROBUST parameter of IMAGR set to 0,
to obtain an optimal compromise between sensitivity and
angular resolution.
In a region of $4' \times 4'$,
we detected a total of 54 sources in one or more of these
individual images; the positions and integrated flux densities of
these sources for each epoch are given in Table 2.
We also list if they have been detected previously in the centimeter
range. 

To obtain a deeper image,
the four individual maps were restored with a beam of 
$0\rlap.{''}32 \times 0\rlap.{''}24$; PA = $0^\circ$,
the largest beam of the four individual maps.
The four maps were then averaged with relative weights corresponding
to the rms of each map.
The resulting image had an rms of 0.03 mJy beam$^{-1}$ and 23 additional
sources were detected above the 5-$\sigma$ level, in the
range of 0.1 to 0.3 mJy. Given its weakness, these
23 additional sources could not have been detected in
previous observations.
The positions and integrated 
flux densities of these sources are given in Table 3.

The position of most of the sources detected is shown in Figure 1, overlapped on
the H$\alpha$ image of O'Dell \& Wong (1996).
Seven of the 77 centimeter sources fall outside of the region shown in Figure 1.
Following Fomalont et al. (2002), we estimate that in a field
of $4' \times 4'$ the \sl a priori \rm number of expected 3.6 cm sources 
above 0.1 mJy is $\sim$0.6. We then conclude that probably
one out of the 77 sources could be a background object, but that we are
justified in assuming that practically all the members of the radio cluster are 
associated with Orion.

\section{OVERALL CHARACTERISTICS OF THE SAMPLE OF RADIO SOURCES}

\subsection{New Centimeter Detections}

Out of the 77 radio sources detected, 39 are new centimeter
detections. These 39 new
centimeter sources are the 16 sources without a GMR
denomination in Table 2 and all 23 sources in
Table 3. Most of the new 39 sources are, as expected, quite
faint. The exceptions are sources 1 and 33 that reach a few
mJy at least in one of the epochs observed. Source 1 is relatively
away from the Orion cluster and was probably detected previously
but simply not reported assuming it is not related to Orion
(see discussion in section 4).  

\subsection{Counterparts at Other Wavelengths}

In Table 4 we list the counterparts of the
radio sources at near-IR, visible and X-ray
wavelengths from the surveys of Ali \& Depoy (1995), O'Dell \& Wen (1994),
and Feigelson et al. (2002), respectively.
Other counterparts, when found in the SIMBAD database,
are listed in the last column.
A counterpart was taken
as such if its position was within 1$''$ of the radio position.
The radio positions are estimated to be accurate to $\sim 0\rlap.{''}05$.
Source numbering was assigned based on increasing right ascension.

Only 9 out of the 77 sources (1, 11, 15, 24, 25, 26, 27, 35,
and 54) do not have a counterpart previously reported in the
literature. These 9 sources have flux densities in the range
of 0.20 to 2.96 mJy.
Their nature is unclear, but we speculate that they may be highly
obscured protostars that remain undetectable even at near-IR
wavelengths.  Centimeter emission, either thermal or non-thermal,
can result from a variety of mechanisms present in low- and high-mass
star forming regions, as discussed by Rodr\'\i guez, G\'omez, \&
Reipurth (2003).  

Of the 77 centimeter sources, 49 have a near-IR counterpart, while 46
are detected in the visible, and 38 in X-rays.  A few of the sources
have reported millimeter counterparts or are associated with H$_2$O
masers (see Table 4).  Because of the orientation of the Orion nebula,
with the main ionization front lying in front of the more distant and
more opaque PDR, any bright proplyds we detect in the radio can be expected to be
visible optically, since they lie in the relatively unobscured ionized
gas. This in turn suggests that radio sources without an optical
counterpart may be non-thermal sources rather than proplyds.  In fact,
of the 19 radio sources without optical counterparts that
were detected in the images from individual epochs, 16 are variable,
further supporting the idea that they are non-thermal.  

\subsection{Systematic Obscuration?}

The sources at the west of the region seem to be systematically more
obscured than those at the east. If we take the first 30 sources in
Table 4, we find that only 8 (27\%) have a visible counterpart, while
if we take the last 47 sources, we find that 38 (81\%) have a visible
counterpart.  This could be attributed to the fact that the main
ionization front of Orion is closer to us in the SW of the nebula and
that more stars could be embedded in the PDR behind the ionization
front in this direction (Wen \& O'Dell 1995).

\subsection{Time Behavior}

In Table 4 we also note if the radio sources are time
variable (above a level of 30\%).
Of the total of 54 sources detected
in one or more of the individual epochs,
36 show time variability (by more than
30\%) between the observed epochs.
The time behavior of the radio sources in Orion
has been studied in detail by Felli et al. (1993a),
where flux densities are reported for 13 epochs at two wavelengths
for observations made over a seven-month span in 1990.
In particular, we confirm the remarkable stability of source 59
(their source 6), that was used by them as a secondary reference
source for flux calibration. This confirmation supports the
reliability of our time monitoring. 

On the other hand, a more detailed comparison of our results
with those of Felli et al. (1993a) reveals new aspects
of the time behavior of the sources. Felli et al. (1993a) firmly
identify 13 objects as constant, thermal sources:
their objects 2, 3, 5, 6, 7, 8, 11, 16, 17, 19, B, I, and K
(corresponding to our sources 71, 65, 61, 59, 52, 48, 42, 53,
60, 74, 12, 19, and 55, respectively). 
We find that 8 out of these 13 sources are indeed non variable
(see Table 4).
There is then an apparent discrepancy in sources 53, 60, 74, 19, and 55
(our nomenclature) in the sense that Felli et al. (1993a)
find them to be constant while we find them to be variable.
However, a revision of Table 2 shows that the
five latter sources have relatively smooth variation over
the several years covered by our study. We then tentatively
conclude that these sources are slowly variable, on a timescale
of years, and thus could not become evident in the Felli et al. (1993a)
study that had a duration of seven months.

Felli et al. (1993a) also firmly classify another 13 objects as
nonthermal, variable, stellar emitters: their sources
4, 9, 10, 12, 13, 15, 23, 25, A, C, F, G, H
(corresponding to our sources 64, 45, 43, 41, 40, 49, 47, 38,
6, 14, 76, 73, and 18, respectively).
We find that 8 out of these 13 sources are indeed variable. 
Again, there is an apparent discrepancy in sources 64, 45, 43, 40,
and 38 (our nomenclature) in the sense that Felli et al. (1993a)
find them to be variable while we find them to be constant. 
We speculate that these sources are a combination of thermal,
constant emission with a nonthermal variable contribution that
during our observations was either off or had constant flux density.

Source 55 (source K in the nomenclature of Felli et al. 1993a)
has been rising steadily in flux density between 1994 and
1997.
It is also interesting that source 56,
located at only
$\sim$1$''$ east of source 55 and first
reported here, also shows a monotonic increase in
its flux density from 1994 to 1997 (see Table 2 and Figure 2).
We speculate that these two sources are being externally
photoionized by a source with time variable ionizing flux
or that the increase in radio flux density of source 55 is 
producing an increase in the ionization of source 56.
In Figure 2 it can be seen that source 56 is bow-shaped and 
roughly points
to the Trapezium region, confirming the proplyd identification
given by Bally et al. (1998).

\subsection{Circular Polarization}

Only one of the sources, number 6, showed clear evidence of circular polarization
at levels of $-$10, $+$5, $+$10, and $-$4 percent for 
1994, 1995, 1996, and 1997,
respectively.
We use here the convention that left-handed circular 
polarization is given as negative, while
right-handed circular polarization is given as positive.

\section{COMMENTS ON SELECTED INDIVIDUAL SOURCES}

In this section we briefly discuss
sources that presented interesting characteristics that
have not been presented in the rest of the paper.

\subsection{Source~1}

This relatively bright source has no reported counterpart in the
literature. Given its flux density of $\sim$3 mJy, the probability of
it being an extragalactic source in the region of $4'\times 4'$ considered
is only $\sim$0.01, suggesting it is associated 
with Orion but it has been missed
in previous studies given its relatively large
displacement from the center of the
cluster.

\subsection{Source~6 = GMR~A}

This bright and time-variable radio source was 
discussed in detail by Garay et al. (1987)
and Felli et al. (1993a) and is known as GMR A. It has near-IR
and X-ray
counterparts but no known visible counterpart
(Felli et al. 1993a). Recently, it presented
a spectacular flare at millimeter and X-ray wavelengths (Bower et al. 2003;
Nakanishi et al. 2003; Getman et al. 2003). Our centimeter data shows
that the source presented a large increase, of about 30
(from 1.31 to 56.6 mJy), between 1996 and 1997.
As noted before, this source shows large percentages of circular polarization
that combined with its large variability indicates that the emission has
a gyrosynchrotron nature.

\subsection{Source~47}

This source was detected by us only in 1996 at the
faint level of 0.85 mJy. However, it is source 23 in
Felli et al. (1993a), where it was found to exceed 20 mJy
at 6 cm in August of 1990. It seems to be then a highly variable
source. 

\subsection{Source~54}

We did not find a reported counterpart for this radio source in the literature.
However, in Figure 1 we can see that there is an optical object associated
with it. The radio morphology of the source (see Fig. 3) suggests it is a
proplyd, with its tail pointing to the north, away from the Trapezium.

\section{PROPLYDS}

Using the table of proplyds 
of O'Dell \& Wen (1994), we find that 26 of the centimeter
sources (numbers 20, 23, 28, 30, 34, 37, 40,
41, 42, 43, 44, 45, 47, 48, 52, 53, 58, 59, 60, 61,
62, 64, 65, 66, 68, and 72) are coincident with objects
identified as proplyds.
In the region shown in Figure 1, O'Dell \& Wen (1994)
report a total of 33 proplyds, of which we detect 26 of them.
The nature of the 7 "optical" proplyds without 
a radio counterpart (objects 139$-$320, 144$-$334,
163$-$322, 168$-$309, 168$-$326N, 171$-$315, and 191$-$350 in
O'Dell \& Wen [1994]) is unclear.

As expected from the externally ionized 
nature of the proplyds, these sources show no or modest (less
than a factor of two) time variability
with the exception of sources 30, 41, 44, 47, and 68.
Source 30 corresponds to the proplyd 154$-$324 of
O'Dell \& Wen (1994), who identify it as a semistellar,
circularly symmetric
proplyd with a star associated. We believe that the radio variability
could be explained if the emission is coming mostly 
from the magnetosphere of the
associated star. 
Source 41 corresponds to the proplyd 158$-$314 of
O'Dell \& Wen (1994), with an unclear morphology due to saturation and
associated with an IR star.  
The radio and optical positions are coincident within a few tenths of
arcsec with the Trapezium O7 star $\theta^1$ Ori A (=V1016 Ori = HD 37020).
This star is well known to be very variable in the radio
(Felli et al. 1991; 1993a).
We then attribute the bright, time-variable radio emission not
to a possibly associated proplyd but to emission associated with the
star $\theta^1$ Ori A.
Source 68 corresponds to the proplyd 173$-$341 of
O'Dell \& Wen (1994), with a semistellar size and 
elongated morphology with diffuse boundary. Again,
the time variable radio emission could be
coming from an active magnetosphere. We have no explanation for the
variability observed in sources 44 and 47.

The morphology seen in the optical for the proplyds is also evident
in the radio image. In Figs. 4 and 5 we show two of the brightest
proplyds: source 59 (corresponding to
proplyd 167$-$317 of O'Dell \& Wen 1994; see
Henney et al. 2002 for a high angular resolution
MERLIN 6 cm image) and source 72 (corresponding to
proplyd 177$-$341 of O'Dell \& Wen 1994). The first object 
shows a thin tail, while the second shows
a wide tail. From hydrodynamical considerations,
Dyson, Hartquist, \& Biro (1993) and
Falle et al. (2002) attribute this difference in morphology
to the speed of the wind impacting on the proplyds. If the wind is supersonic,
the tail occupies a sizable opening angle, while if the wind is subsonic
(presumably due to mass loading in the trajectory between the star and
the proplyd), the tail is thin.
Cant\'o et al. (1998) have included the effects of the
shadowing of photoionizing radiation by the proplyds, concluding that
the appearance of the tails is also influenced by
whether or not the shadowed
region is optically thick to the scattered Lyman continuum radiation emanating
from recombinations of H~II within the surrounding nebula.
This model was used by O'Dell (2000) to explain his Orion observations. 

We finally note that sources 32, 56, 63, and 69 in the northeastern part
of Figure 1 have a bow-shaped morphology pointing roughly to the Trapezium
region. Indeed, all four sources are identified by Bally et al.
(1998) as proplyds. 
An image of source 56 is shown in Fig. 2, while source 63 is shown in Fig.
6. 

\section{SOURCES IN OMC-1S}

A region in Orion that has received considerable attention in recent
years is OMC-1S. Located about 1$'$ to the SW of the Trapezium, it is
about one-tenth as bright as the BN-KL region in the 1.3 mm thermal
radiation of dust (Mezger, Zylka, \& Wink 1990).  Remarkably, several
optical outflows
(HH~202, HH~269, HH~529, HH~203, HH~204, and possibly HH~528) 
seem to emanate from a region only a few arcseconds across.
O'Dell \& Doi (2003) refer to this region as the Optical Outflow
Source (OOS), that they locate at $\alpha(2000) = 05^h~ 35^m~ 14\rlap.^s56;
\delta(2000) = -05^\circ~ 23'~ 54\rlap.{''}0$.
The best candidate for the source
of this remarkable outflow activity is the heavily embedded object
TPSC-1 (Lada et al. 2000), which is seen only in the near-infrared L
(=3.5 $\mu$m) band.

Within $\sim 30{''}$ of the OOS there are reported two centers of 
CO outflows.
A high velocity molecular outflow of about 100 km s$^{-1}$ expels
material to the NW (blueshifted) and SE (redshifted) (Rodr\'\i guez-Franco 
et al. 1999), and the blueshifted lobe is associated
with the HH~625 optical flow.
The center of this high velocity CO outflow is located at
$\alpha(2000) = 05^h~ 35^m~ 13\rlap.^s5;
\delta(2000) = -05^\circ~ 23'~ 53{''}$.
A more extended, low-velocity (5 km~s$^{-1}$)
bipolar molecular outflow is oriented NE (blueshifted)-SW (redshifted)
(Schmid-Burgk et al. 1990). The center of this CO outflow seems to
coincide with the 1.3 mm continuum source FIR4 (Mezger et al. 1990),
located at $\alpha(2000) = 05^h~ 35^m~ 13\rlap.^s4;
\delta(2000) = -05^\circ~ 24'~ 13{''}$. 
 
In particular, no centimeter sources have been reported in 
association with any of these three centers of outflow activity.
Our deep 3.6 cm image of Orion reveals the presence of two faint sources 
(13 and 15)
located near (at $\sim 7{''}$ and $4{''}$, respectively) of the OOS,
but not exactly coincident
with it (see Fig. 7). We also detect for the
first time a cm source (source 10, see Fig. 8) that is coincident within
the millimeter positional error
($\sim 2{''}$) with FIR4, the presumed exciting source of the
low-velocity CO outflow. Since we only have flux densities for 
these sources at one wavelength (3.6 cm), we cannot advance significantly
in the interpretation of their nature.

\section{A FILAMENTARY RADIO STRUCTURE NEAR OMC-1S}

Of the four data sets used, only the 1995 observations were made
with short spacings that could allow the proper imaging of
extended ($\geq 2{''}$) structures. An image made with this
data set alone shows the presence of an elongated structure
near OMC-1S (see Fig. 9). The radio morphology of this source is
suggestive of a radio jet, however it could be as well a sharp
ionization front. The optical image (see Fig. 9) favors  
the latter interpretation since there is practically no emission on
the southern side of the structure (presumably the neutral region), while
there is extended emission to the north.
The filament appears to be part of a larger, spur-like
structure, as seen in the 2 cm, C-configuration VLA image
of Felli et al. (1993b; see their Fig. 6).
The total 3.6 cm flux density of this structure inside
a region of $6\rlap.{''}3 \times 1\rlap.{''}9$ is $\sim$20 mJy.
We have used a calibrated H$\beta$ image of Orion
(O'Dell \& Doi 1999; O'Dell, Peimbert, \& Peimbert 2003) to
derive that the H$\beta$ flux inside the same region, after correcting
for the extended background, is $\sim 4.5 \times 10^{-12}$ ergs cm$^{-2}$
s$^{-1}$. The extinction correction at this position is
$C(H \beta)$ = 0.58 (O'Dell et al. 2003), which implies
an extinction-corrected flux of
$\sim 1.7 \times 10^{-11}$ ergs cm$^{-2}$
s$^{-1}$.
Assuming that the emission originates in ionized hydrogen with an electron
temperature of 10$^4$ K that is optically thin in the radio, and
following Caplan \& Deharveng (1986) we find that the H$\beta$
flux implies a 3.6 cm flux density of $\sim$45 mJy, about a factor of
2 higher than the value measured by us. This discrepancy could be due 
to an overestimation of the extinction correction, to the radio emission
being optically-thick, or to the gas being unusually cold ($\sim$3000 K). 
A similar discrepancy has been found in the comparison of radio and
optical data of some Herbig-Haro objects (Pravdo et al. 1985; Anglada et al.
1992). None of the explanations seems attractive. The extinction
is smoothly changing in this
region and our estimate should be adequate. The 
peak temperature brightness of the
filament is modest, $T_B \simeq$ 150 K, suggesting that we are observing
optically-thin free-free emission. We also considered the contrast-decrease
effect caused by the ambient H~II region (Pastor, Cant\'o, \& Rodr\'\i guez
1991; Henney, Garc\'\i a-D\'\i az, \& Kurtz 2001), but 
the average free-free opacity of 
Orion at 3.6 cm is $\sim$0.1 and this effect is too small to
explain the discrepancy.

\section{CONCLUSIONS}

We have presented the analysis of sensitive, high angular 
resolution ($0\rlap.{''}3$) VLA
observations at 3.6 cm toward the Orion region of
recent star formation.
Our main two conclusions are summarized below:

1) In a region of $4' \times 4'$ around the Orion Trapezium we 
detect a total of 77 compact 3.6 cm sources, of which 39 are
new centimeter detections. We briefly discuss
the characteristics of these sources
and of their counterparts at other bands.

2) In the OMC-1S region we detect  
three faint sources that may be related to the sources that
power the multiple outflows that 
emanate from this region.

\vspace{0.5cm}

\acknowledgments

We thank Marcello Felli for his detailed comments
that significantly improved our paper.
LFR and SEK acknowledge the support
of DGAPA, UNAM, and of CONACyT (M\'exico).
This research has made extensive use of the SIMBAD database, 
operated at CDS, Strasbourg, France.

\vfill\eject

\clearpage

\begin{figure}
\plotone{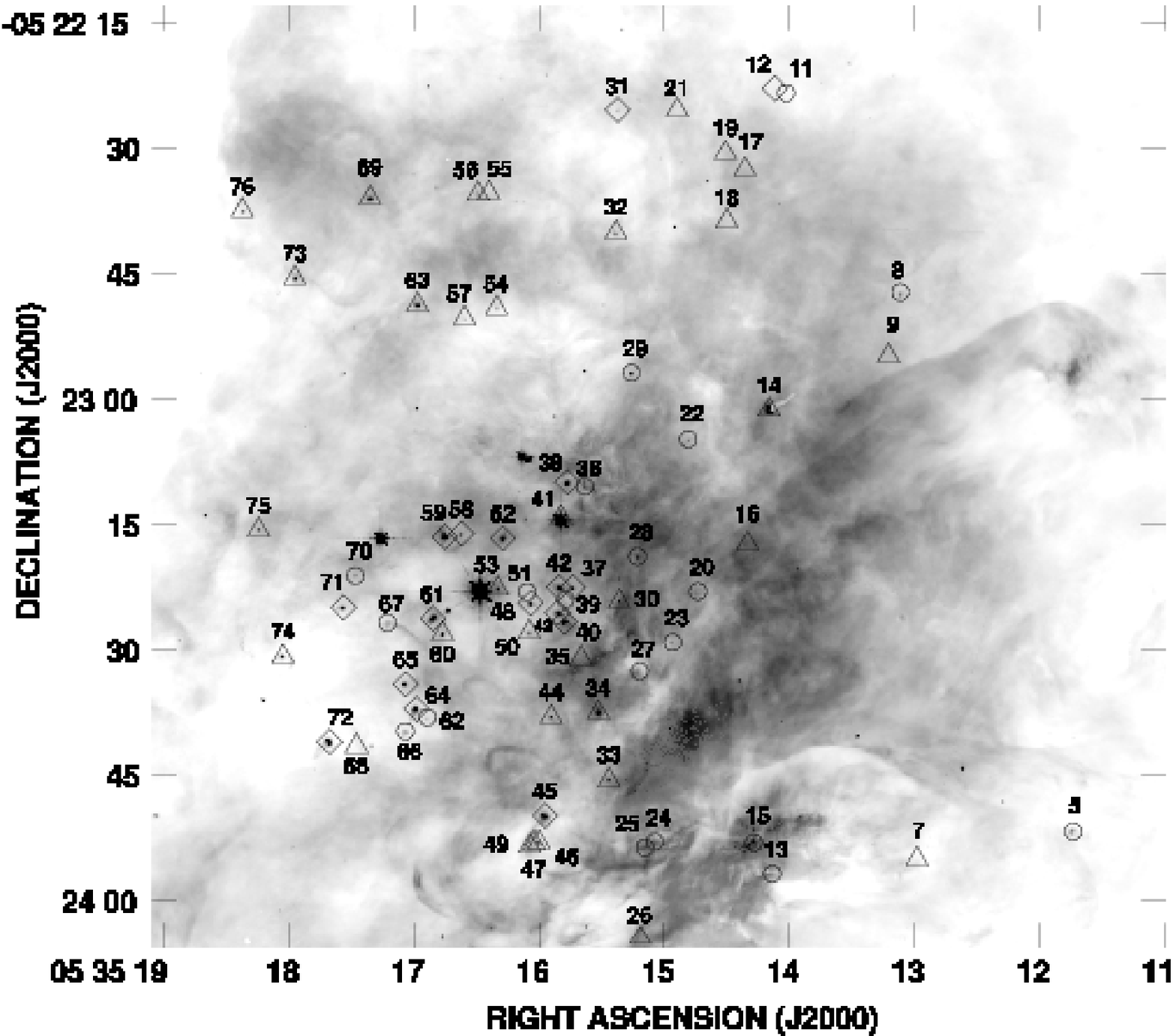}
\caption{Positions of most of the VLA 3.6 cm sources, superposed on
the H$\alpha$ image of O'Dell \& Wong (1996).
Sources 1, 2, 3, 4, 6, 10, and 77 fall outside the region shown. 
The triangles ($\vartriangle$)
indicate time-variable sources, the diamonds ($\Diamond$) indicate non variable
sources, and the circles ($\circ$) indicate faint
sources detected in the averaged image (see Table 3).
The positions of the radio sources is at the center of
the symbols.
\label{fig1}}
\end{figure}

\begin{figure}
\plotone{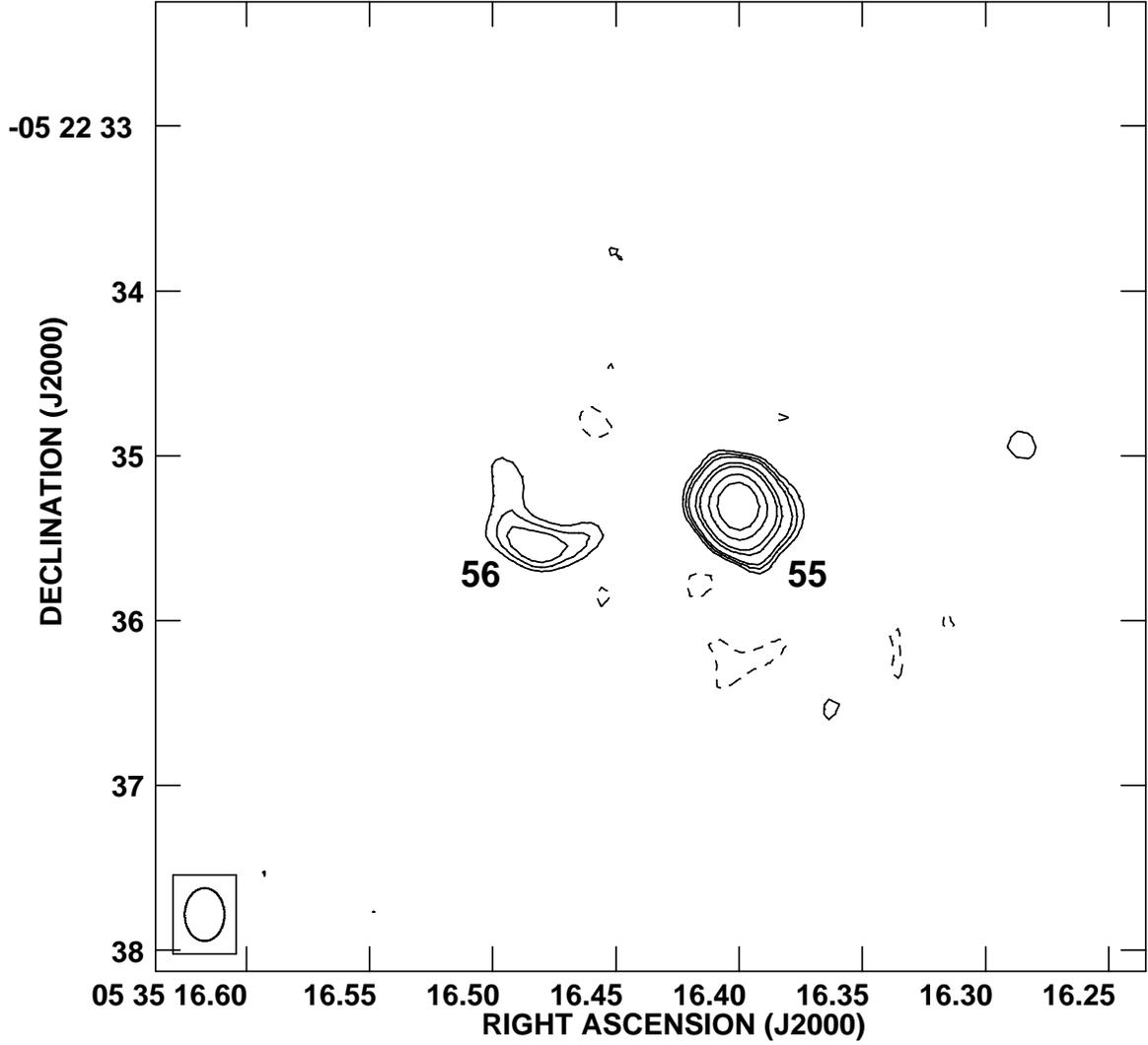}
\caption{VLA contour image at 3.6 cm of the sources 55 and 56,
made from the average of all data.
The contours are -4, -3, 3, 4, 5, 8, 10, 15, 20, and 30
times 36 $\mu$Jy beam$^{-1}$.
The half power contour of the restoring beam
is shown in the bottom left.
\label{fig2}}
\end{figure}

\begin{figure}
\plotone{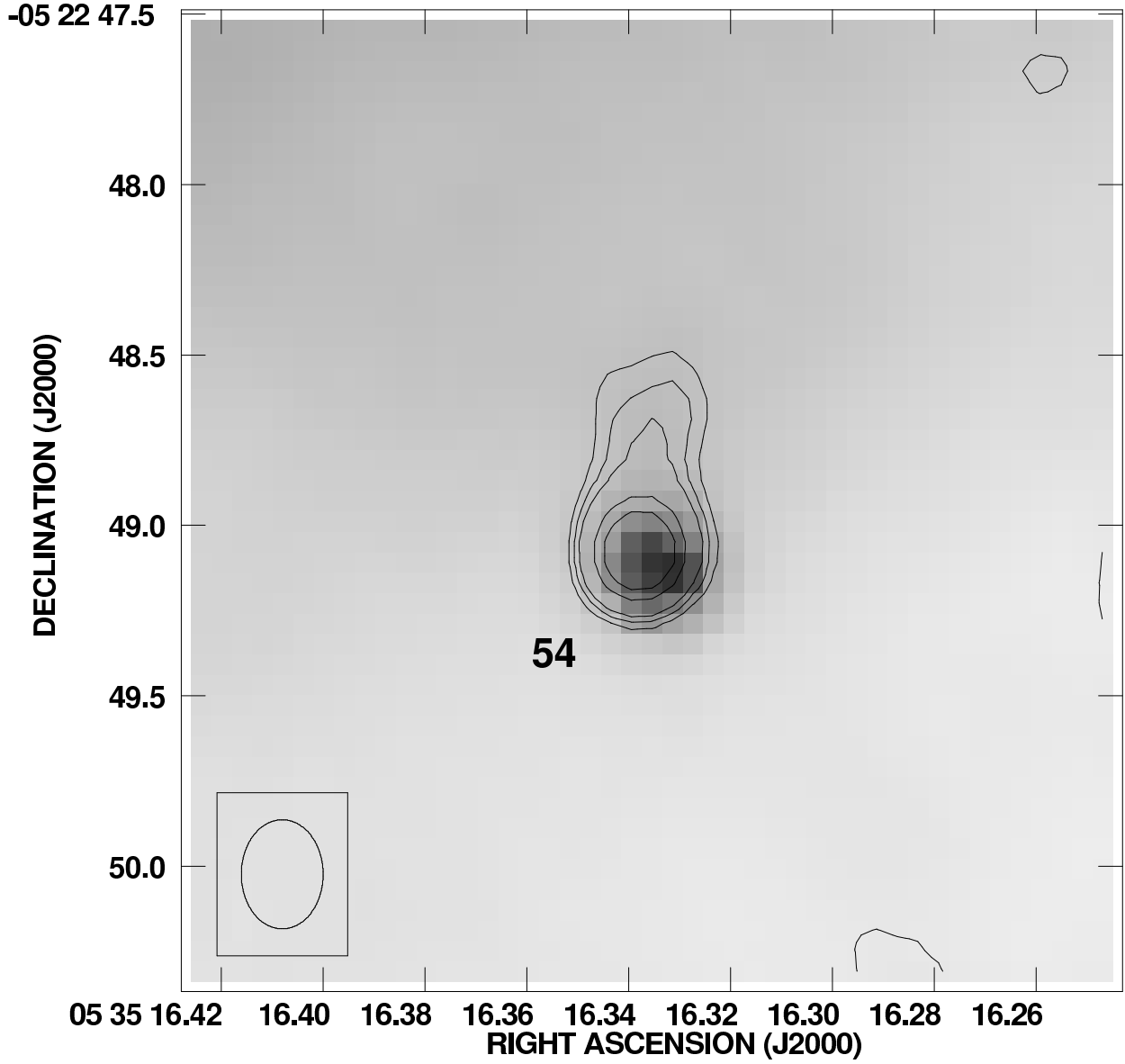}
\caption{VLA contour image at 3.6 cm of the source 54,
superposed on 
the H$\alpha$ image of O'Dell \& Wong (1996), shown in greyscale.
The contours are 3, 4, 5, 8, and 10 
times 35 $\mu$Jy beam$^{-1}$.
\label{fig3}}
\end{figure}

\begin{figure}
\plotone{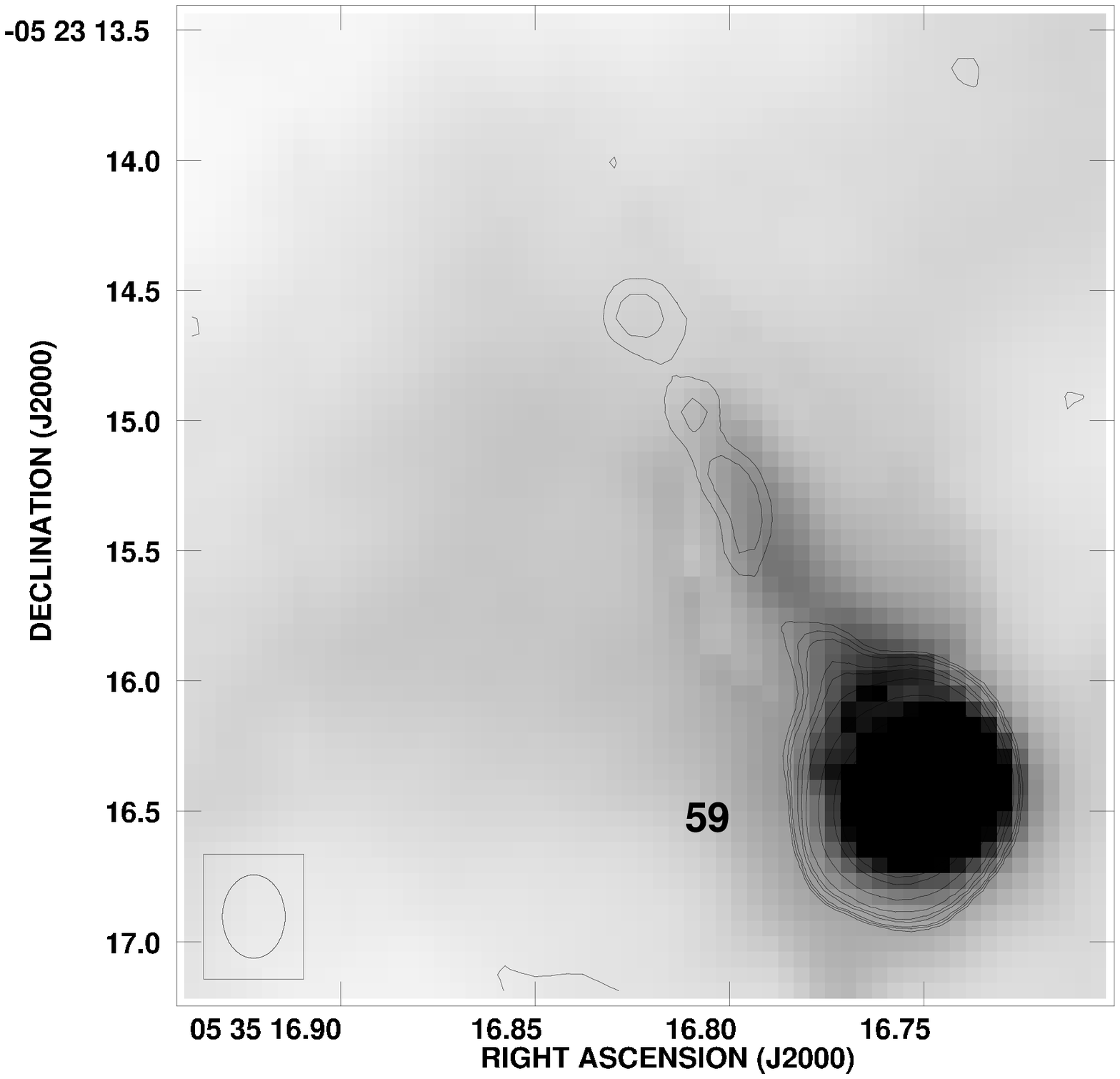}
\caption{VLA contour image at 3.6 cm of the source 59,
superposed on
the H$\alpha$ image of O'Dell \& Wong (1996), shown in greyscale.
The contours are 3, 4, 5, 8, 10, 15, 30 and 60
times 35 $\mu$Jy beam$^{-1}$.
This is an example of a thin tail proplyd.
\label{fig4}}
\end{figure}

\begin{figure}
\plotone{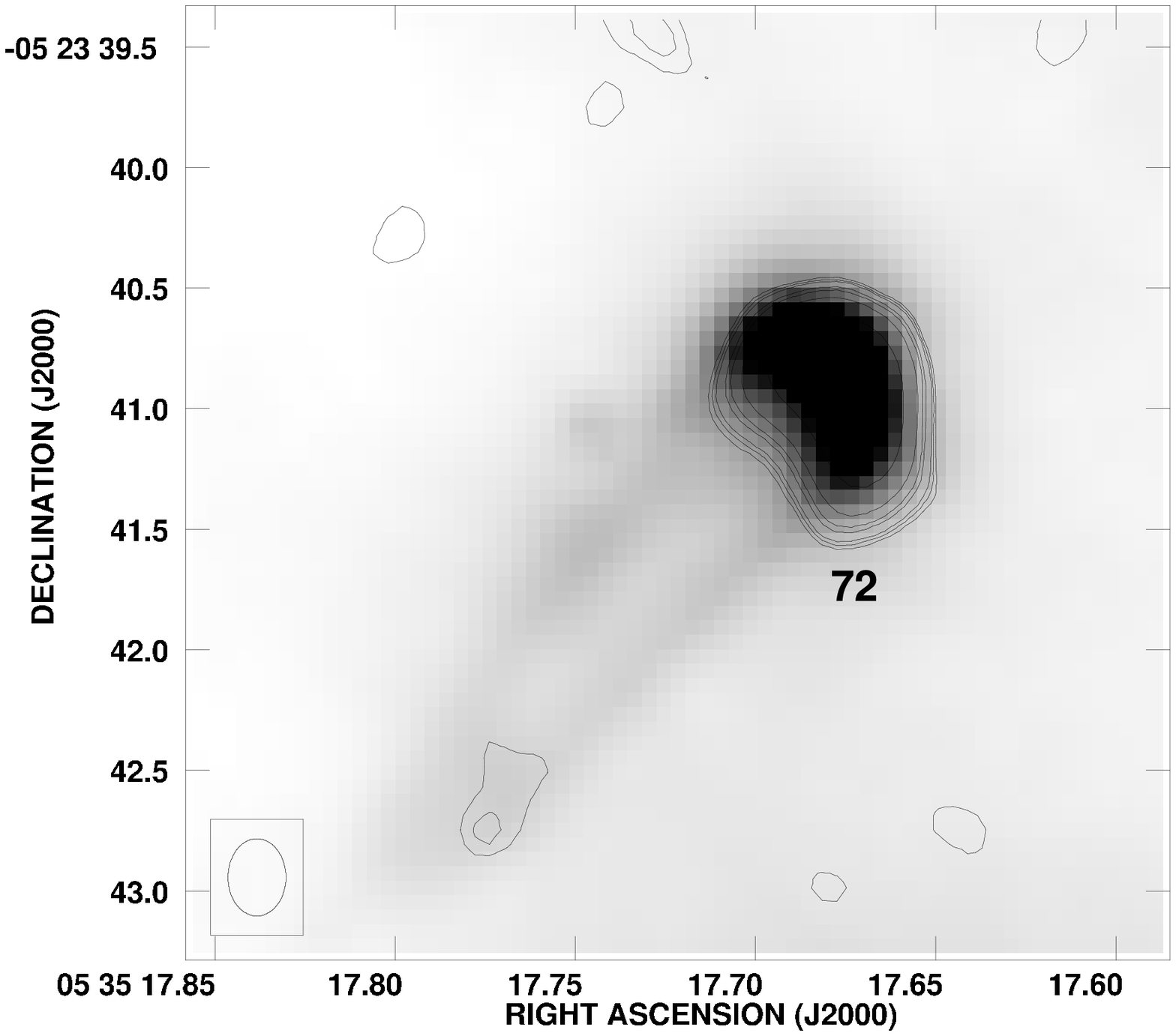}
\caption{VLA contour image at 3.6 cm of the source 72,
superposed on
the H$\alpha$ image of O'Dell \& Wong (1996), shown in greyscale.
The contours are 3, 4, 5, 8, 10, 15, and 30 
times 35 $\mu$Jy beam$^{-1}$.
This is an example of a wide tail proplyd.
\label{fig5}}
\end{figure}

\begin{figure}
\plotone{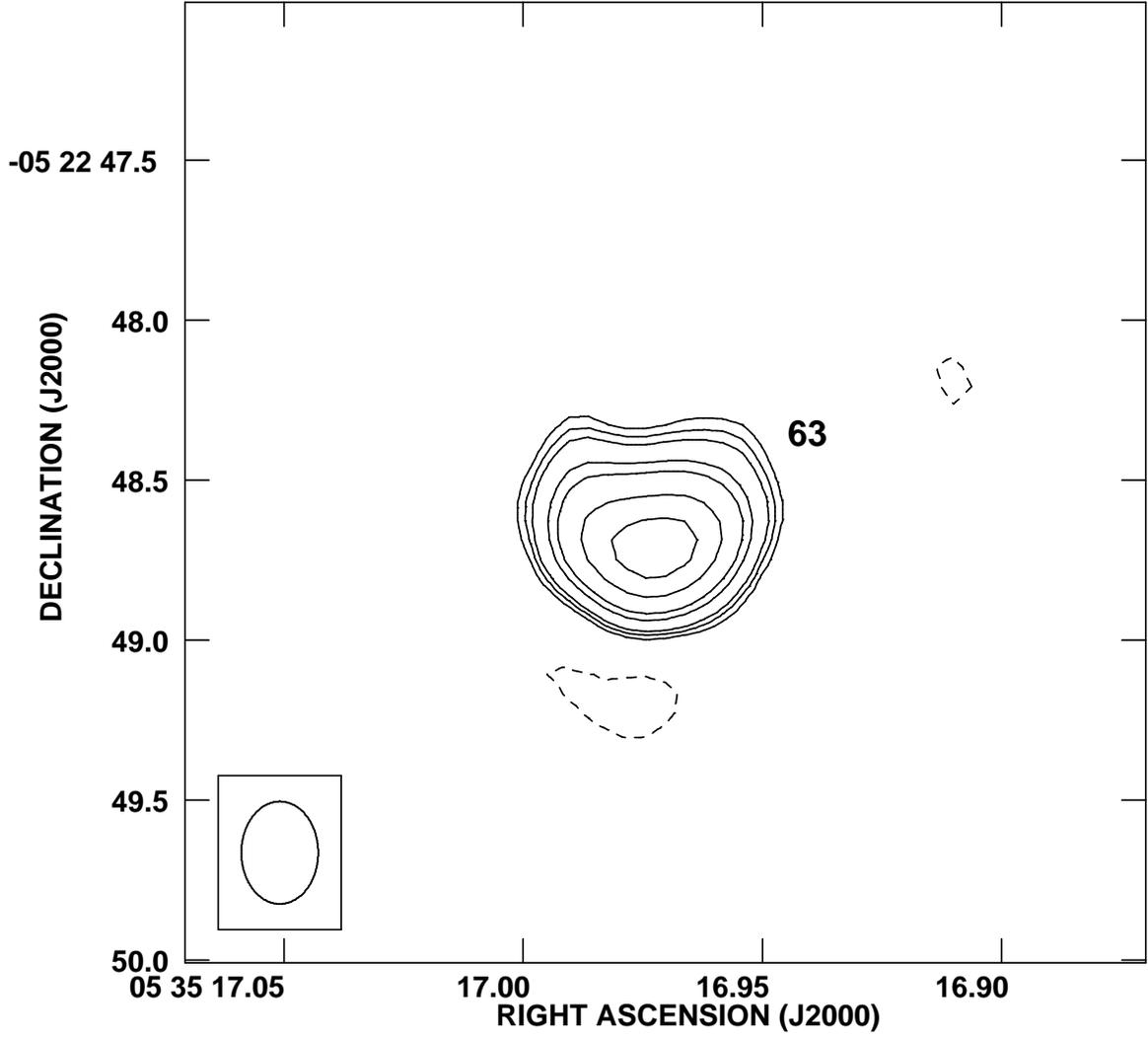}
\caption{VLA continuum image at 3.6 cm of source 63.
The contours are -4, -3, 3, 4, 5, 8, 10, 15, and 20 
times 45 $\mu$Jy beam$^{-1}$.
The half power contour of the restoring beam
is shown in the bottom left.
\label{fig6}}
\end{figure}

\begin{figure}
\plotone{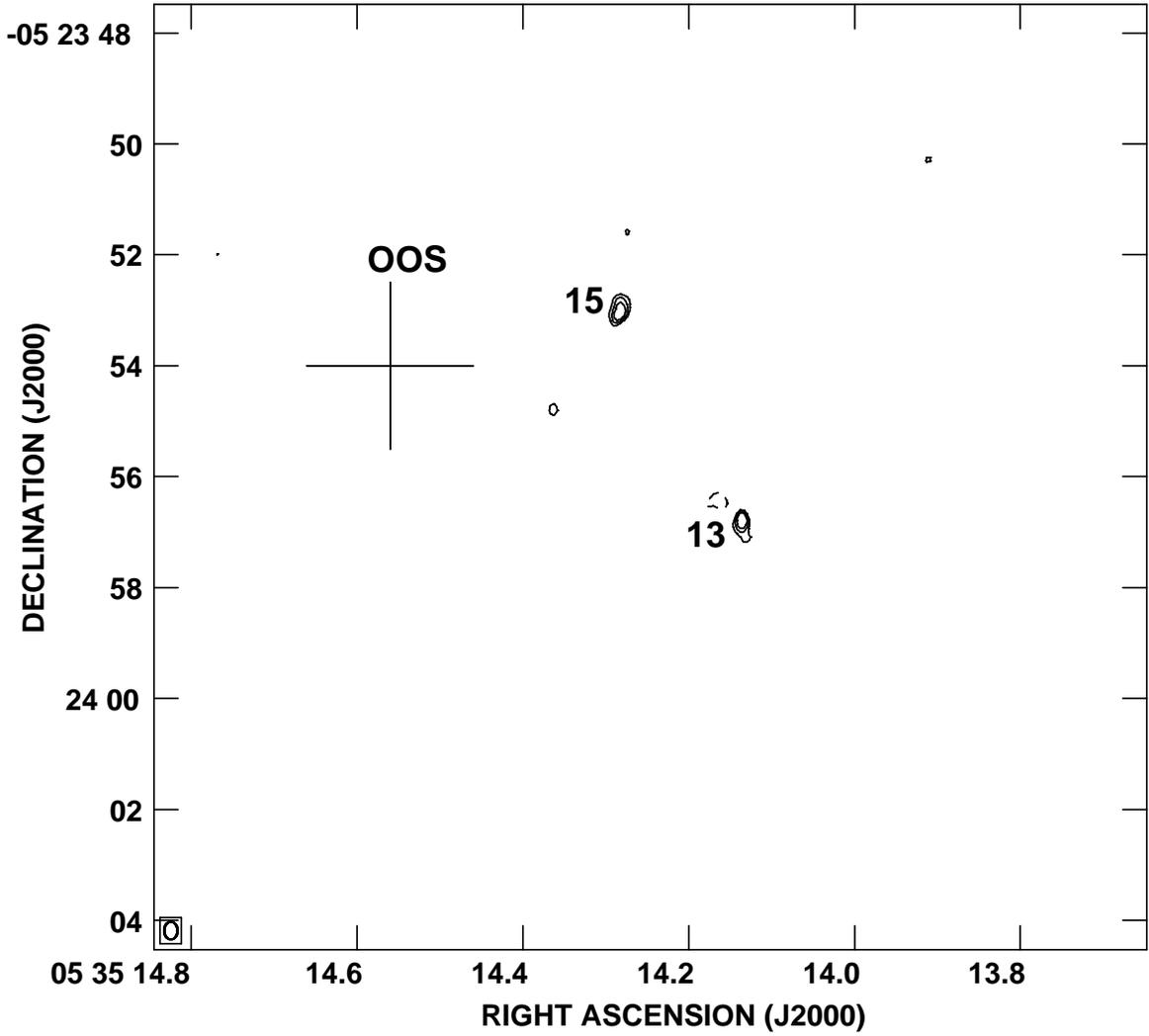}
\caption{VLA continuum image at 3.6 cm of the OOS region.
The faint sources 13 and 15 are first reported here.
The position of the OOS (O'Dell \& Doi 2003)
is indicated with a cross.
The contours are -4, 4, 5, and 6
times 40 $\mu$Jy beam$^{-1}$, the rms noise of this region
of the image.
The half power contour of the restoring beam
is shown in the bottom left.
\label{fig7}}
\end{figure}

\begin{figure}
\plotone{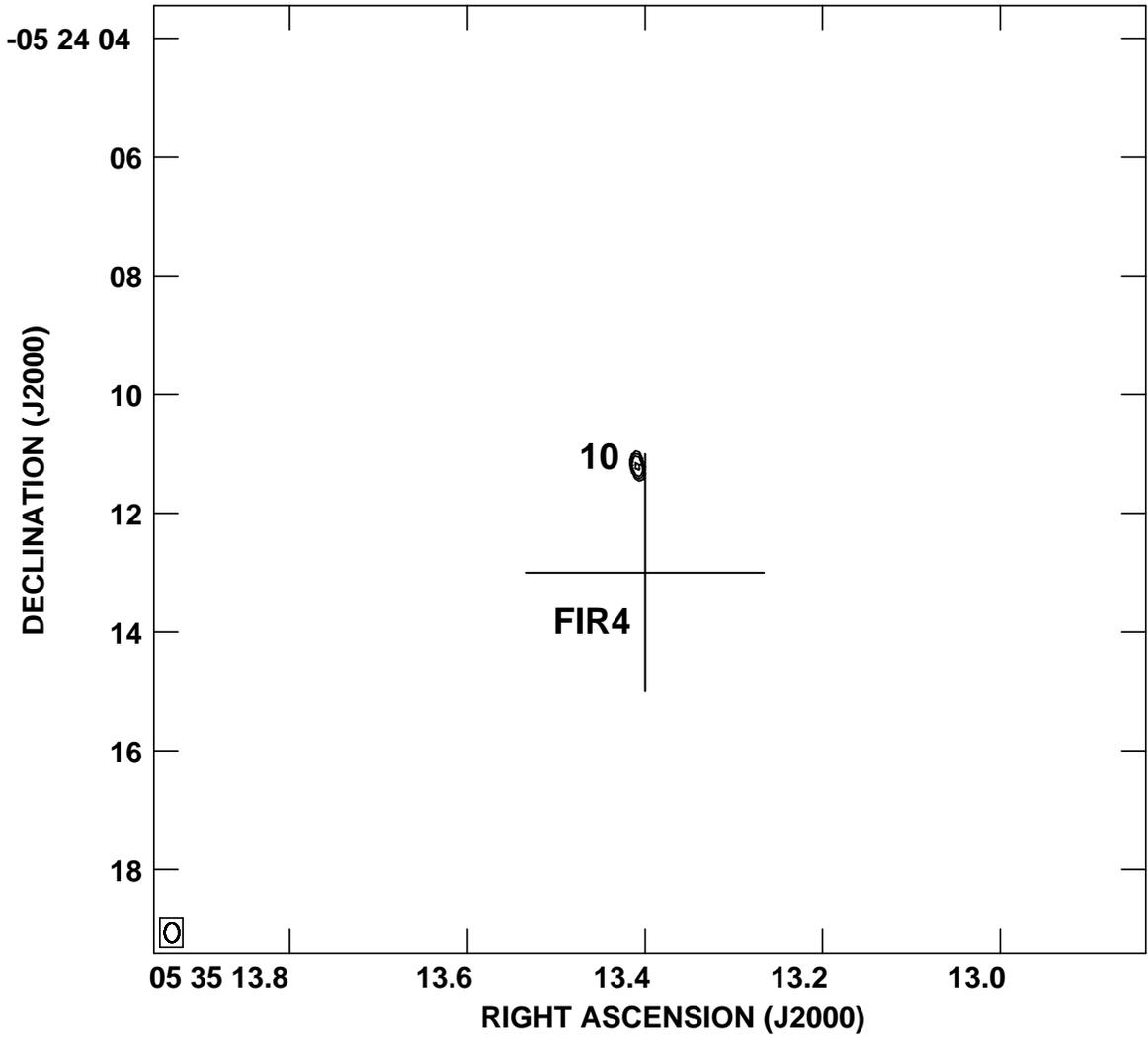}
\caption{VLA continuum image at 3.6 cm of the FIR4 region.
The faint source 10 is first reported here.
The position of FIR4 (Mezger et al. 1990) is indicated with a cross.
The contours are -4, 4, 5, 6, and 8
times 40 $\mu$Jy beam$^{-1}$, the rms noise of this region
of the image.
The half power contour of the restoring beam
is shown in the bottom left.
\label{fig8}}
\end{figure}

\begin{figure}
\plotone{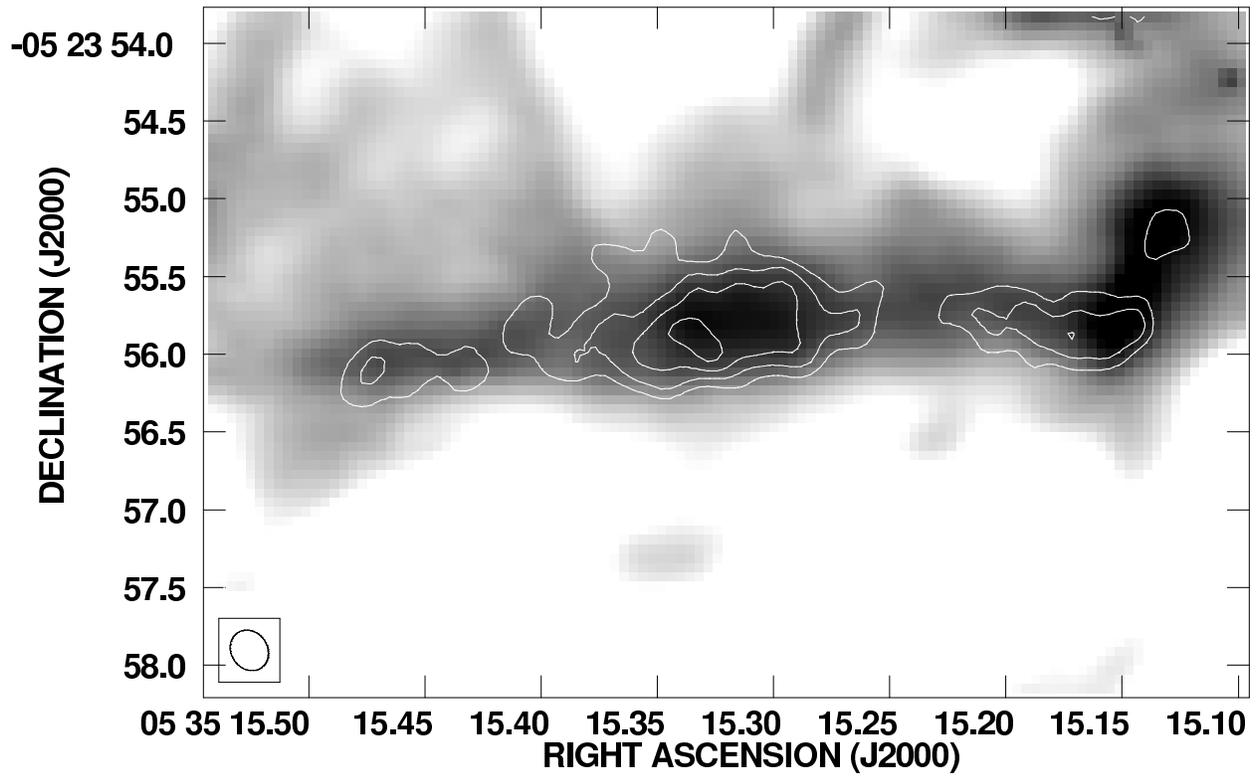}
\caption{VLA contour image at 3.6 cm of the filamentary structure
near OMC-1S, superposed on
the H$\alpha$ image of O'Dell \& Wong (1996), shown in greyscale.
The contours are -3, 3, 4, 5, and 6
times 84 $\mu$Jy beam$^{-1}$.
The half power contour of the restoring beam
is shown in the bottom left.
\label{fig9}}
\end{figure}

\clearpage

\vfill\eject

\begin{table}[htbp]
\small
  \caption{Parameters of the observations at 3.6~cm}
  \begin{center}
    \begin{tabular}{lcccccc}\hline\hline
&\multicolumn{2}{c}{Phase Center} &Phase  &Bootstrapped Flux 
&rms noise & Number of \\
\cline{2-3} 
Epoch &  $\alpha$(J2000) & $\delta$(J2000) & Calibrator & Density (Jy) & 
(mJy) & Antennas \\ 
\hline
1994& 05 35 14.479 & -05 22 30.57 & 0501-019 & 2.46$\pm$0.02 & 0.04
& 18 \\
1995& 05 35 16.019 & -05 22 59.57 & 0541-056 & 0.92$\pm$0.03 & 0.08
& 26 \\
1996& 05 35 14.479 & -05 22 30.57 & 0501-019 & 1.55$\pm$0.01 & 0.05
& 14 \\
1997& 05 35 14.479 & -05 22 30.57 & 0501-019 & 1.54$\pm$0.01 & 0.08
& 14 \\

\hline\hline
    \label{tab:1}
    \end{tabular}
  \end{center}
\end{table}

\vfill\eject

\begin{deluxetable}{l c c c c c c c}
\tabletypesize{\small}
\tablecolumns{8} 
\tablewidth{0pc} 
\tablecaption{Parameters of the 3.6 cm VLA Sources Detected in One or More Individual Images}
\tablehead{
\colhead{}                              &                  
\colhead{}                              &
\colhead{}                              &
\colhead{}                              &
\multicolumn{4}{c}{Flux Density$^a$}   \\ 
\cline{5-8}                             
\colhead{Source}                        & 
\colhead{GMR$^b$}                           &
\colhead{$\alpha_{2000}$}               &
\colhead{$\delta_{2000}$}               &
\colhead{1994}                          &
\colhead{1995}                          &
\colhead{1996}                          &
\colhead{1997}                           
}
\startdata
 1   &   & 05 35 04.855 & -05 23 02.65 &  2.90$\pm$ 0.06   &  2.96 $\pm$ 0.07  &  2.78$\pm$ 0.07    
 &   2.70 $\pm$ 0.08  \\
 2   &   & 05 35 09.764 & -05 21 28.22 &  1.51$\pm$ 0.04   &  $<$ 0.3          &   0.1$\pm$0.03     
 &   $<$0.1           \\
 6   & A & 05 35 11.800 & -05 21 49.20 &  7.68$\pm$ 0.2    &  1.78 $\pm$  0.06  &  1.31 $\pm$ 0.03  
 &   56.6 $\pm$ 0.05  \\
 7   &   & 05 35 12.980 & -05 23 54.95 &  0.30$\pm$ 0.05   &  $<$ 0.3          &  1.08 $\pm$ 0.06   
 &   0.96 $\pm$ 0.07   \\
 9   & Q & 05 35 13.208 & -05 22 54.78 &  0.79$\pm$ 0.05   &  $<$ 0.3          &  0.48 $\pm$ 0.03   
 &   0.17 $\pm$ 0.03    \\
 12  & B & 05 35 14.115 & -05 22 22.85 &  3.73$\pm$ 0.04   &  3.36 $\pm$ 0.3   &  3.43 $\pm$ 0.1    
 &   3.90 $\pm$ 0.07    \\
 14  & C & 05 35 14.160 & -05 23 01.29 &  5.55$\pm$ 0.07   &  2.57 $\pm$ 0.1   &  5.77 $\pm$ 0.1    
 &   5.98 $\pm$ 0.1     \\
 16  &   & 05 35 14.333 & -05 23 17.29 &  0.31$\pm$ 0.05   &  0.57 $\pm$ 0.1   &  0.31 $\pm$ 0.06   
 &   0.12 $\pm$ 0.04    \\
 17  & L & 05 35 14.355 & -05 22 32.52 &  0.71$\pm$ 0.05   &  1.07 $\pm$ 0.09  &  1.09 $\pm$ 0.07   
 &   1.60 $\pm$ 0.07    \\
 18  & H & 05 35 14.499 & -05 22 38.67 &  0.41$\pm$ 0.03   &  0.17 $\pm$ 0.05  &  0.77 $\pm$ 0.07   
 &   1.31 $\pm$ 0.06   \\      
 19  & I & 05 35 14.510 & -05 22 30.53 &  0.43$\pm$ 0.04   &  0.45 $\pm$ 0.04  &  0.83 $\pm$ 0.07   
 &   0.87 $\pm$ 0.05   \\
 21  & D & 05 35 14.896 & -05 22 25.39 &  0.76$\pm$ 0.04   &  0.52 $\pm$ 0.06  &  2.04 $\pm$ 0.07   
 &   0.84 $\pm$ 0.05  \\
 26  &   & 05 35 15.181 & -05 24 03.61 &    $<$ 0.3        &  1.25 $\pm$ 0.2   &  1.74 
$\pm$ 0.1    
 &   1.09 $\pm$ 0.1   \\
 30  &   & 05 35 15.359 & -05 23 24.16 &  0.56$\pm$ 0.05   &  0.70 $\pm$ 0.06  &  0.38 $\pm$ 0.08   
 &   0.19 $\pm$ 0.05   \\
 31  &   & 05 35 15.372 & -05 22 25.39 &  0.12$\pm$ 0.05   &  0.22 $\pm$ 0.06  &  0.30 $\pm$ 0.08   
 &   0.14 $\pm$ 0.05   \\
 32  &   & 05 35 15.385 & -05 22 40.05 &  0.49$\pm$ 0.04   &  0.96 $\pm$ 0.05  &  0.97 $\pm$ 0.07   
 &   0.53 $\pm$ 0.04    \\
 33  &   & 05 35 15.440 & -05 23 45.55 &  3.20$\pm$ 0.1    &  0.29 $\pm$ 0.07  &  4.10 $\pm$ 0.1    
 &   4.01 $\pm$ 0.1   \\
 34  &14 & 05 35 15.525 & -05 23 37.48 &  4.81$\pm$ 0.1    &  5.81 $\pm$ 0.1   &  6.19 $\pm$ 0.2    
 &   5.84 $\pm$ 0.02  \\
 35  &   & 05 35 15.632 & -05 23 31.51 &  $<$0.3           &  0.20 $\pm$ 0.05  &        
$<$0.3 
 &   0.97 $\pm$ 0.02  \\
 37  &26 & 05 35 15.730 & -05 23 22.52 &  1.47$\pm$ 0.06   &  1.48 $\pm$ 0.04  &  1.40 $\pm$ 0.09   
 &   1.42 $\pm$ 0.07  \\
 38  &25 & 05 35 15.773 & -05 23 09.92 &  4.14$\pm$ 0.06   &  5.29 $\pm$ 0.05  &  5.52 $\pm$ 0.1    
 &   4.89 $\pm$ 0.07  \\
 40  &13 & 05 35 15.798 & -05 23 26.61 &  9.45$\pm$ 0.1    &  9.68 $\pm$ 0.1   &  10.0 $\pm$ 0.1    
 &   8.73 $\pm$ 0.04 \\
 41  &12 & 05 35 15.824 & -05 23 14.15 &  5.64$\pm$ 0.07   &  9.57 $\pm$ 0.09  &  9.53 $\pm$ 0.1    
 &   23.6 $\pm$ 0.09 \\
 42  &11 & 05 35 15.840 & -05 23 22.51 &  10.6$\pm$ 0.09   &  10.8 $\pm$ 0.1   &  11.0 $\pm$ 0.1    
 &   10.9 $\pm$ 0.1 \\
 43  &10 & 05 35 15.851 & -05 23 25.54 &  3.69$\pm$ 0.09   &  4.56 $\pm$ 0.1   &  3.70 $\pm$ 0.1    
 &   4.25 $\pm$ 0.1 \\
 44  &24 & 05 35 15.902 & -05 23 38.01 &  0.71$\pm$ 0.05   &  1.80 $\pm$ 0.1   &  1.23 $\pm$ 0.1    
 &   1.59 $\pm$ 0.09 \\
 45  & 9  & 05 35 15.951 & -05 23 49.82 &  5.50$\pm$ 0.1    &  6.85 $\pm$ 0.2   &  5.99 $\pm$ 0.1    
 &   5.78 $\pm$ 0.1 \\
 46  &  & 05 35 16.001 & -05 23 53.00 &  0.98$\pm$ 0.1    &  1.96 $\pm$ 0.1   &  1.09 $\pm$ 0.1    
 &   2.78 $\pm$ 0.1 \\
 47  &23 & 05 35 16.039 & -05 23 53.12 &  $<$ 0.2          &  $<$ 0.3          &  0.85 $\pm$ 0.1    
 &   $<$ 0.28    \\
 48  &8  & 05 35 16.068 & -05 23 24.35 &  4.69$\pm$ 0.07   &  4.68 $\pm$ 0.1   &  4.26 $\pm$ 0.1    
 &   4.03$\pm$ 0.09 \\
 49  &15 & 05 35 16.073 & -05 23 07.12 &  3.91$\pm$ 0.08   &  4.34 $\pm$ 0.1   &  3.64 $\pm$ 0.1    
 &   5.51$\pm$ 0.09  \\


 50 & 22 & 05 35 16.078 & -05 23 27.86 &  1.23$\pm$ 0.09  &  1.98 $\pm$ 0.1   &  2.18 $\pm$ 0.1  &  
3.16$\pm$ 0.1  \\
 52 & 7  & 05 35 16.290 & -05 23 16.62 &  9.60$\pm$ 0.08  &  9.17 $\pm$ 0.1   &  10.7 $\pm$ 0.1  &  
11.0$\pm$ 0.1  \\
 53 & 16 & 05 35 16.329 & -05 23 22.66 &  2.63$\pm$ 0.06  &  2.96 $\pm$ 0.1   &  3.38 $\pm$ 0.1  &  
3.58$\pm$ 0.09  \\
 54 &    & 05 35 16.339 & -05 22 49.10 &  0.64$\pm$ 0.03  &  0.56 $\pm$ 0.08  &  1.00 $\pm$ 0.06 &  
1.90 $\pm$ 0.01  \\
 55 & K  & 05 35 16.401 & -05 22 35.32 &  1.21$\pm$ 0.05  &  1.85 $\pm$ 0.09  &  2.47 $\pm$ 0.08 &  
4.19 $\pm$ 0.08  \\
 56 &    & 05 35 16.488 & -05 22 35.42 &  0.28$\pm$ 0.09  &  0.29 $\pm$ 0.08  &  0.82 $\pm$ 0.1  &  
1.21 $\pm$ 0.1  \\
 57 &    & 05 35 16.594 & -05 22 50.30 &  0.18$\pm$ 0.09  &  0.36 $\pm$ 0.08  &  0.54 $\pm$ 0.09 &  
0.53 $\pm$ 0.09 \\
 58 & 21 & 05 35 16.620 & -05 23 16.07 &  1.76$\pm$ 0.07  &  1.58 $\pm$ 0.07  &  1.59 $\pm$ 0.1  &  
1.38 $\pm$ 0.01 \\
 59 & 6  & 05 35 16.753 & -05 23 16.47 &  22.0$\pm$ 0.09  &  22.2 $\pm$ 0.1   &  22.7 $\pm$ 0.1  &  
22.0 $\pm$ 0.1 \\
 60 & 17 & 05 35 16.770 & -05 23 28.06 &  2.62$\pm$ 0.06  &  3.38 $\pm$ 0.08  &  2.97 $\pm$ 0.4  &  
3.52 $\pm$ 0.1 \\
 61 & 5  & 05 35 16.849 & -05 23 26.26 &  16.1$\pm$ 0.09  &  15.6 $\pm$ 0.1   &  16.1 $\pm$ 0.1  
 & 15.7 $\pm$ 0.1 \\
 63 & E  & 05 35 16.971 & -05 22 48.70 &  1.55$\pm$ 0.07  &  2.48 $\pm$ 0.1   &  1.78 $\pm$ 0.3  &  
2.46 $\pm$ 0.4 \\
 64 & 4  & 05 35 16.981 & -05 23 37.01 &  8.16$\pm$ 0.07  &  7.81 $\pm$ 0.1   &  7.08 $\pm$ 0.8  &  
7.56 $\pm$ 0.1 \\
 65 & 3  & 05 35 17.068 & -05 23 34.09 &  3.57$\pm$ 0.07  &  3.87 $\pm$ 0.1   &  3.81 $\pm$ 0.7  &  
4.01 $\pm$ 0.07 \\
 68 &    & 05 35 17.331 & -05 23 41.42 &  0.63$\pm$ 0.1   &  0.75 $\pm$ 0.04  &  1.30 $\pm$ 0.1  &  
1.40 $\pm$ 0.1 \\
 69 & L  & 05 35 17.350 & -05 22 35.92 &  1.67$\pm$ 0.07  &  2.45 $\pm$ 0.1   &  1.74 $\pm$ 0.1 
 &  2.24 $\pm$ 0.07 \\
 71 & 2  & 05 35 17.562 & -05 23 24.89 &  3.60$\pm$ 0.07  &  4.44 $\pm$ 0.1   &  3.86 $\pm$ 0.1   & 
 4.32 $\pm$ 0.07 \\
 72 & 1  & 05 35 17.674 & -05 23 40.97 &  9.63$\pm$ 0.1   &  8.11 $\pm$ 0.1   &  7.43 $\pm$ 0.1   & 
 9.63 $\pm$ 0.1 \\
 73 & G  & 05 35 17.951 & -05 22 45.50 &  4.48$\pm$ 0.07  &  0.58 $\pm$ 0.1   &  0.77 $\pm$ 0.06  &
 4.10 $\pm$ 0.1 \\
 74 & 19 & 05 35 18.047 & -05 23 30.77 &  3.50$\pm$ 0.08  &  2.93 $\pm$ 0.09  &  3.92 $\pm$ 0.08  &
4.25 $\pm$ 0.1 \\
 75 &    & 05 35 18.241 & -05 23 15.62 &  $<$ 0.2         &  0.33 $\pm$ 0.07  &  0.43 $\pm$ 0.04  &
0.53 $\pm$ 0.1 \\
 76 & F  & 05 35 18.373 & -05 22 37.44 &  18.0$\pm$ 0.07  & 12.21 $\pm$ 0.05  &  9.44 $\pm$ 0.1  &  
20.5 $\pm$ 0.07 \\
 77 & V  & 05 35 21.058 & -05 23 49.10 &  2.93$\pm$ 0.08  &  0.38 $\pm$ 0.06  &  4.28 $\pm$ 0.1  &  
3.66 $\pm$ 0.07 \\

\tablecomments{
                (a): Total flux density corrected for primary beam response
                The uncertainty in the flux densities was determined from
                measurements of a region free of sources near each source.\\
                (b): GMR referes to the centimeter sources detected by
                Garay, Moran, \& Reid (1987), Garay (1987), 
                Churchwell et al. (1987),
                Garay (1989), 
                and Felli et al. (1993a; 1993b).
\\
}           
\enddata
\end{deluxetable}
\vfill\eject

\begin{deluxetable}{l c c c}
\tablecaption{Parameters of the 3.6 cm VLA Sources Detected in the Averaged Image}
\tablehead{
\colhead{Source}                        & 
\colhead{$\alpha_{2000}$}               &
\colhead{$\delta_{2000}$}               &
\colhead{Flux Density}              
}
\startdata
3  &   05 35 10.481 & -05 22 45.69  &   0.16$\pm$0.02  \\
4  &   05 35 10.734 & -05 23 44.71  &   0.30$\pm$0.02  \\
5  &   05 35 11.733 & -05 23 51.71  &   0.24$\pm$0.02  \\  
8  &   05 35 13.113 & -05 22 47.20  &   0.11$\pm$0.02  \\  
10 &   05 35 13.409 & -05 24 11.21  &   0.33$\pm$0.03  \\
11 &   05 35 14.026 & -05 22 23.39  &   0.29$\pm$0.03  \\
13 &   05 35 14.136 & -05 23 56.77  &   0.32$\pm$0.04 \\   
15 &   05 35 14.283 & -05 23 53.05  &   0.30$\pm$0.03  \\
20 &   05 35 14.730 & -05 23 23.01  &   0.29$\pm$0.04  \\
22 &   05 35 14.808 & -05 23 04.78  &   0.33$\pm$0.03  \\
23 &   05 35 14.932 & -05 23 29.08  &   0.32$\pm$0.04  \\  
24 &   05 35 15.075 & -05 23 52.96  &   0.26$\pm$0.05  \\  
25 &   05 35 15.154 & -05 23 53.63  &   0.31$\pm$0.05  \\ 
27 &   05 35 15.195 & -05 23 32.55  &   0.30$\pm$0.05  \\
28 &   05 35 15.213 & -05 23 18.82  &   0.34$\pm$0.04  \\
29 &   05 35 15.262 & -05 22 56.86  &   0.24$\pm$0.03  \\
36 &   05 35 15.635 & -05 23 10.41  &   0.37$\pm$0.05  \\
39 &   05 35 15.789 & -05 23 23.96  &   0.28$\pm$0.04  \\
51 &   05 35 16.096 & -05 23 23.05  &   0.30$\pm$0.04  \\
62 &   05 35 16.891 & -05 23 38.05  &   0.38$\pm$0.04  \\ 
66 &   05 35 17.069 & -05 23 39.72  &   0.32$\pm$0.06  \\
67 &   05 35 17.211 & -05 23 26.73  &   0.28$\pm$0.04  \\ 
70 &   05 35 17.467 & -05 23 21.11  &   0.32$\pm$0.04  \\ 

\enddata
\end{deluxetable}
\vfill\eject

\begin{deluxetable}{l c c c c c c }
\tabletypesize{\small}
\tablecolumns{6} 
\tablewidth{0pc} 
\tablecaption{Variability and Counterparts of the 3.6 cm VLA Sources}
\tablehead{
\colhead{}                              &                  
\colhead{}                              &
\colhead{}                              &
\multicolumn{4}{c}{Counterpart$^a$}   \\ 
\cline{4-7}                             
\colhead{}                       &
\colhead{Position-based}                       &
\colhead{Time}                    &
\colhead{near-IR}               &
\colhead{visible}               &
\colhead{X-rays }                & 
\colhead{}                    \\
\colhead{Source}                       & 
\colhead{Names$^b$}             &
\colhead{Variable?}                    &
\colhead{(AD95)}               &
\colhead{(OW94)}               &
\colhead{(F02)}                 &
\colhead{other}                           
}
\startdata
 1 & 049-303  & N   & --   & --      &  --   & --   \\
 2 & 098-128  & Y   & -- & 98-128  &  244  & [HC00] 621;[H97b] 336   \\
 3 & 105-246  & --  & -- & --  &  262  & --   \\
 4 & 107-345  & --  & -- & --      &  270   & --   \\
 5 & 117-352  & --  & -- & 117-352 &  --   & --   \\
 6 & 118-149  & Y   & 3108 & --      &  297  & --   \\
 7 & 130-355  & Y   & --   & --      &  337  & --   \\
 8 & 131-247  & --  & --   & 131-247 &  --   & [HC00] 707;[H97b] 9031   \\
 9 & 132-255  & Y   & 2717 & --      &  351  & --      \\
 10 & 134-411 & --  & --   & --      &  --.  & [GWV98]  \\
 11 & 140-223 & --  & --   & --      &  --   & -- \\
 12 & 141-223 & N   & --   & --      & 399   & [CHS01] 8470;[LRY00] 44\\
 13 & 141-357 & --  & --   & --      &  --   & [GWV98]    \\
 14 & 142-301 & Y   & -- & -- &  --   & [SCB99] 23 \\
 15 & 143-353 & --  & --   &  --     &  --   & -- \\
 16 & 143-317 & Y  & --   &   --    & 406   & -- \\
 17 & 144-233 & Y  & --   &   --    & 407   & [SCB99] 32;[LRY00] 50 \\
 18 & 145-239 & Y   & --   &   --    & 416   & --   \\
 19 & 145-231 & Y   & --   &   --    & --    & [D93] IRc2 A;[GWV98]\\
 20 & 147-323 & --  & -- & 147-323 & 432    & [H97b] 451;[HC00] 302 \\
 21 & 149-225 & Y  &  --  & --      & 442   & [H97b] 472b\\
 22 & 148-305 & --   & --  & 148-305 & --    & [H97b] 9073 \\
 23 & 149-329 & --  & --   & 149-329 & 445   & --   \\
 24 & 151-353 & --  & --   & --      & --    & --  \\
 25 & 152-354 & --   & --   & --      & --    & --   \\
 26 & 152-404 & Y   & --   & --      & --    & --   \\
 27 & 152-333 & --  & --   & --      & --    & --    \\
 28 & 152-319 & --  & 3512 & 152-319 & 459    & [HC00] 318    \\
 29 & 153-257 & --  & 3257 & --      & 461   & --  \\
 30 & 154-324 & Y  & -- & 154-324 & --    & [BSD98] 19 \\
 31 & 154-225 & N   &  --  & 154-225 & 465    & [H97b] 9096;[HC00] 475;[SCB99] 74 \\


32 & 154-240 & Y  &  --  & -- & --     & [LRY00] 103 \\
33 & 154-346 & Y  & --  & 154-346 & 474    & [HB97b] 475;[MLL95] J053515-052346 \\
34 & 155-337 & Y  & 3143 & 155-338 & --    & [BSD98] 23\\
35 & 156-332 & Y  &  --  & --      &  --    & --    \\
36 & 156-310 & --  & -- &  --     &  --    & EQ J053515-052311      \\
37 & 157-323 & N & 3253 & 157-323 & --    & [BSD98] 14  \\
38 & 158-310 & N  & --   & 158-310 &  495   & [HC00] 342;[SCB99] 100 \\
39 & 158-324 & --  & 3254   &  --     &  --    & --   \\
40 & 158-327 & N & --   & 158-327 &  --    & [BSD98] 13;[HC00] 287;[H97b] 489\\
41 & 158-314 & Y  & --   & 158-314 & 498    & [HC00] 336;[LRY00] 121  \\
42 & 158-323 & N  & --   & 158-323 & 501    & [MLL95] J053515-052322;[H97b] 488a   \\
43 & 158-326 & N  & --  & 158-326 &  --    & [BSD98] 12 \\
44 & 159-338 & Y  & --   & 159-338 &  --    & [BSD98] 20;[HC00] 250\\
45 & 160-350 & N  & -- & 159-350  &  508   & [SCB99] 104  \\
46 & 160-353W & Y  & --   &  --     &  513   & [SCB99] 106 \\
47 & 160-353E & Y  & --   & 160-353 &  --    & [H97b] 503; [SCH01] 61  \\
48 & 161-324 & N  & --   & 161-324 &  --   & [HC00] 296 \\
49 & 161-307 & Y  & --   & 161-307 &  519   & [H97b] 1863a\\
50 & 161-328 & Y  & --   & 161-328 &  --    & [BSD98] 9;[HC00] 285;[LRY00] 135\\
51 & 161-323 & --  & --   & 161-323 &  517    & [LRY00] 138   \\
52 & 163-317 & N & --   & 163-317 &  529   & [BSD98] 6;[HC00] 322    \\
53 & 163-323 & Y  & --   & 163-322 & --    & [BSD98] 1;[SCB99] 127  \\
54 & 163-249 & Y  & --   & --      &   --   &  --   \\
55 & 164-235 & Y  & 3518 & --      &  --   &  --     \\
56 & 165-235 & Y  & --   & 165-235 &  543    & [H97b] 519b   \\
57 & 166-250  & Y  & --   & 166-250 &  --    & [LRY00] 164 \\
58 & 166-316 & N  & -- & 166-316 & 549    & [BSD98] 7\\  
59 & 168-316 & N  & --   & 167-317 &  554    & [MLL95] J053516-052316;[HC00] 323\\
60 & 168-328 & Y  & --   & 168-328 &  555   & [HC00] 284;[BSD98] 5  \\


61 & 168-326 & N  & --   & 168-326S  & --   & [MLL95] J053516-052326;[SCB99] 149   \\
62 & 169-338 & --  & --   & 169-338   & --   & [BSD98] 17 \\
63 & 170-249 & Y & 3260  & 170-249   & 571   & [BSD98] 31  \\
64 & 170-337 & N  & --   & 170-337   & 570  & [H97b] 534    \\
65 & 171-334 & N  & --   & 171-334   & 576  & [H97b] 538;[BSD98] 15\\
66 & 171-340 & --  & 3144 & 171-340   & 574   & --    \\
67 & 172-327 & -- & --   & 172-327   & --   & --    \\
68 & 173-341 & Y & --   & 173-341    & --   & --    \\
69 & 174-236 & Y  & --   & 173-236   & --   & [H97b] 548;[HC00] 440 \\
70 & 175-321 & --  & --   & 175-321   & 596   & [H97b] 9180    \\
71 & 176-325 & N & 3529 & --        & --   & --    \\
72 & 177-341 & N  & --   & 177-341   & --  & [H97b] 558;[BSD98] 26;   \\
73 & 180-246 & Y  & 3168 & 179-246   & 626  & --  \\
74 & 180-331 & Y  & --   & 180-331   & --   & [H97b] 9211;[BSD98] 25;[LRY00] 234\\
75 & 182-316 & Y  & --   & --        & 646  & --    \\
76 & 184-237 & Y  & --   & 183-238   & 651  & [H97b] 589;[HC00] 436    \\
77 & 211-349 & Y  & 3140 & 210-349   & 760  & [H97b] 669   \\

\tablecomments{
                (a): AD95 = Ali \& Depoy 1995 (K-band);
                BPN78 = Beckwith 1978(K band); 
                BSD98 = Bally et al. 1998 (UV);
                CHS01 = Carpenter et al. 2001 (K-band);
                D93  =  Dougados et al. (1993) (L-band);
                EQ = Tsunemi et al. 2001 (X-rays);
                F02 = Feigelson et al. 2002 (X-rays)
                GWV98 = Gaume et al. 1998 (water masers);
                HAB = Hyland et al. 84 (K band); 
                Hb97 = Hillenbrand 1997 (I-band);
                HC00 = Hillenbrand \& Carpenter 2000 (K-band);
                LRY00 = Luhman et al. 2000 (K-band);
                MLL95 = Mundy et al. 1995 (radio millimeter);
                OW94 = O'Dell \& Wen 1994 (visible);
                SCH01 = Schulz et al. 2001 (X-rays);
                SCB99 = Simon et al. 1999 (K-band).}
\tablecomments{
                (b): After the convention of O'Dell \& Wen (1994).
\\
}           
  
\enddata
\end{deluxetable}

\vfill\eject

\end{document}